\documentclass[]{aa}
\usepackage{graphicx}
\usepackage{txfonts}
\begin{document}

   \title{Terrestrial planet formation during giant planet formation and giant planet migration I: The first 5 million years}

   \author{R. Brasser\inst{1,2}\fnmsep\thanks{rbrasser@konkoly.hu}}
   \institute{Konkoly Observatory, HUN-REN CSFK, MTA Centre of Excellence; Konkoly Thege Miklos St. 15-17, H-1121 Budapest, Hungary 
        \and
        Centre for Planetary Habitability (PHAB), University of Oslo; Sem Saelands Vei 2A, N-0371 Oslo, Norway}
\authorrunning{Brasser}
   \date{}

\abstract
{Terrestrial planet formation (TPF) is a difficult problem that has vexed researchers for decades. Numerical models are only partially successful at reproducing the orbital architecture of the inner planets, but have generally not considered the effect of the growth of the giant planets. Cosmochemical experiments suggest that the nucleosynthetic isotopic composition of bodies from beyond Jupiter is different from that of the inner solar system. This difference could have implications for the composition of the terrestrial planets.}
{I aim to compute how much material from the formation region of the gas giants ends up being implanted in the inner solar system due to gas drag from the protoplanetary disc, how this implantation alters the feedstocks of the terrestrial planets, and whether this implantation scenario is consistent with predictions from cosmochemistry.}
{I dynamically model TPF as the gas giants Jupiter and Saturn are growing using the Graphics Processing Unit (GPU) software Gravitational ENcounters with GPU Acceleration (GENGA). The evolution of the masses, radii and orbital elements of the gas giants are precomputed and read and interpolated within GENGA. The terrestrial planets are formed by planetesimal accretion from tens of thousands of self-gravitating planetesimals spread between 0.5 au and 8.5 au. The total mass of the inner planetesimal disc and outer disc are typically 2 and 3 Earth masses ($M_\oplus$) respectively, and the composition of the planetesimals changes from non-carbonaceous-like to carbonaceous-like at a prescribed distance, ranging from 2 au to 5 au.} 
{Here I report on the first 5~Myr of evolution. At this time approximately 10\% to 25\% of the mass of planetesimals in the Jupiter-Saturn region is implanted in the inner solar system, which is more than what cosmochemical models predict; this amount can be reduced by reducing the total mass of the outer planetesimal disc, and extrapolation from the results suggest a mass of 1~$M_\oplus$ would suffice. The implantation initially sets up a composition gradient in the inner solar system, with the fraction of outer solar system material increasing with increasing distance to the Sun. The planetesimals that remain in the inner solar system have a mixed composition, which ultimately could have implications for late accretion.}
{The growth of the gas giants scatters planetesimals in their vicinity into the inner solar system, which changes the isotopic composition of the terrestrial planets. The planetesimal disc in the vicinity of the gas giants may not have been very massive, around 1~$M_\oplus$. The inner planetesimal disc may not have extended much farther than 2 au otherwise embryos do not grow fast enough to produce Mars analogues. This could mean that the region of the current asteroid belt never contained much mass to begin with. The implantation scenario could also explain the existence of active asteroids in the main belt.}


\maketitle
%
\section{Introduction}
Dynamical modelling of terrestrial planet formation (TPF) has a long history, dating back to the early works of \citet{WetherillStewart1989}, and the state of the art N-body simulations of \citet{KI1995,KI1996,KI1998} using the GRAvity PipE (GRAPE) specialised hardware \citep{Ito1990}. To date there is an overabundance of dynamical models of TPF in existence. Past and present dynamical models either rely on planetesimal accretion, or pebble accretion, which is a mode of planetary growth developed by \citep{OK2010,LJ2012}. \\

The first model that relied on planetesimal accretion to grain traction is dubbed the `classical model' \citep{Chambers2001,Raymond2006,Raymond2009,OBrien2006,Woo2021}, wherein the giant planets were assumed to have fully formed and reside on their current orbits. The terrestrial planets formed by planetesimal accretion from a planetesimal disc roughly situated between 0.5 au and 4 au with a surface density slope simiar to the minimum-mass solar nebula \citep{Hayashi1981}. A variation on this model assumes Jupiter and Saturn on circular orbits \citep{Raymond2009}. The growth of the planets roughly proceeded as follows: as the planets grew and planetesimals were depleted, the initial runaway mode of growth made way to an oligarchic growth mode \citep{KI1998}, wherein a convoy of lunar to Mars mass planetary embryos at their isolation mass \citep{Lissauer1987} stir up and accrete the remaining planetesimals while the embryos remain roughly evenly spaced apart. The planetesimals can provide dynamical friction to the embryos, keeping their eccentricities and inclinations low \citep{OBrien2006}. Eventually this configuration becomes dynamically unstable, and the embryos undergo a phase of mutual collisions -- dubbed the giant impact phase -- and they merge to form Venus and Earth \citep{Chambers2001}.\\

Unfortunately, the classical model suffered from the `small Mars' problem, in which the mass of formed planets near Mars' current orbit were systematically too massive \citep{Chambers2001,OBrien2006,Raymond2006,Raymond2009,Woo2021}. Proposed solutions for this problem are: limiting the width of the initial planetesimal disc in the annulus model \citep{Hansen2009}, invoking the migration of the giant planets to truncate the planetesimal disc with the Grand Tack model \citep{Walsh2011}, lowering the planetesimal surface density beyond Mars with the depleted disc model \citep{Izidoro2014,MahBrasser2021}, steepening the surface density slope of the planetesimal disc \citep{Izidoro2015}, and invoking an early dynamical instability of the giant planets to stir up the inner Solar System \citep{Clement2018,Clement2019}. Each of these models have their pros and cons in terms of reproducing the mass-semimajor axis distribution, Angular Momentum Deficit \citep[AMD;][]{Laskar1997}, growth rates of Earth and Venus \citep{Kleine2009,Rudge2010}, and timing of the Moon-forming event \citep{Jacobson2014}. A detailed review of TPF can be found in \citet{Morbidelli2012}, and a brief discussion of each model's outcomes are given in \citet{Lammer2021}.\\

The great diversity of planetesimal accretion models in existence betrays the fact that further progress relying on simplified dynamical models running on single Central Processing Unit (CPU) cores is limited. Simulations beginning with a few thousand planetesimals and planetary embryos take a few months to complete, and higher planetesimal numbers are prohibitively expensive in terms of required computing resources. This has led the creation of the proprietary Lagrangian Integrator for Planetary Accretion and Dynamics (LIPAD) \citep{Levison2012}, which uses tracer particles to follow the collisional/accretional/dynamical evolution of a large number of kilometer-sized planetesimals through the entire growth process of becoming planets. Collisions are treated in a statistical manner and the code is parallelised, using multiple CPU cores. A study of TPF with LIPAD indicates that the small Mars problem could have partially been an artefact of the numerical limitations of a single CPU core, initial conditions used in earlier models, as well as embryo growth times exceeding 5~Myr beyond 1.5~au \citep{WL2019}. As such, LIPAD offered promising solutions to study TPF. The increasing popularity of computing on graphics cards (Graphical Processing Unit, or GPU) and availability of hardware and dedicated N-body codes that run on GPUs such as Gravitational ENcounters with GPU Acceleration (GENGA) \citep{grimmstadel2014,grimmetal2022}, opened up additional pathways of exploration beyond LIPAD: it became possible to simulate tens of thousands of self-gravitating planetesimals on one GPU card, and thereby to track the growth of Mars from planetesimal accretion \citep{Woo2021} as was done by \citet{WL2019}. With GPU computing the small Mars problem did not entirely disappear, indicating that treating collisions may have a strong effect on the outcome \citep{WL2019}. Yet this technological development by itself does not solve all the problems that the existing models have. What is worse, and therefore important, is that understanding TPF requires more than statistically reproducing the masses and orbital properties of the terrestrial planets because the planets also have different compositions. A good model of TPF should be able to reproduce these differences.\\

Apart from the great variety of outcomes resulting from dynamical studies, problems also arise from (mass-independent) isotopic anomalies measured in samples from the Earth, Moon, Mars, Vesta, and various meteorite classes. These anomalies consistently indicate that intrinsic compositional differences existed in various regions of the Sun’s protoplanetary disc \citep{Warren2011}. Experimental analyses indicate that Mars' isotopic composition is consistently different from the Earth in terms of nucleosynthetic anomalies in Ti, Cr, Fe, Zn, Mo and O \citep[see review by][]{Mezger2020}. When correlating these isotopic anomalies of different elements in the planets and meteorites against each other, two major groupings emerge: a carbonaceous or {\it jovian} group from the outer Solar System, and a non-carbonaceous or {\it terrestrial} group from the inner region \citep{BM2020}. In other words, the isotopic composition of Solar System samples show a {\it dichotomy} \citep[e.g.][]{Kruijer2020}. There is debate whether these differences arose from a compositional gradient that existed in the Sun’s protoplanetary disc \citep{Yamakawa2010}. Monte Carlo models that mix various chondrites as isotopic proxies for the building blocks of the Earth and Mars show that these planets formed primarily from {\it terrestrial} group material \citep{Fitoussi2016,Dauphas2017,Dauphas2024}. This reservoir is the likely source of Earth’s water \citep{Piani2020}. These mixing models also consistently indicate that the isotopic compositions of the Earth and Mars require them both to have accreted a few weight percent (wt.\%) of {\it jovian} group material \citep{Fitoussi2016,Dauphas2017,Dauphas2024}. Despite the quality of the measurements that provide the input for these mixing models and their numerical robustness, their conclusions are difficult to explain dynamically: the impact probability of {\it jovian} group material with the terrestrial planets is extremely low if it arrived directly from beyond Jupiter \citep{Brasser2020}.\\

In addition to their different nucleosynthetic isotopic compositions, the Earth and Mars also have disparate growth times. Mars is thought to have almost fully formed some 5 million years (Myr) after the formation of the Solar System \citep{DauphasPourmand2011,TangDauphas2014}, i.e. while the protosolar disc was still around \citep{Wang2017}. The giant planets, by their nature, must also have formed while the protosolar disc was around, and their N isotopes indicate their gaseous envelopes were likely accreted from the protosolar disc \citep{Marty2011,FuriMarty2015}. From Hf-W dating of iron meteorites and arguing that Jupiter is responsible for the dichotomy, \citet{Kruijer2017} infer that Jupiter must have formed in 1-2 Myr after the formation of the Solar System, i.e. contemporary with Mars. In contrast, the Earth took much longer, from 10 to 50 Myr \citep[e.g.][]{Yin2002,Kleine2009,YuJacobsen2011}, culminating in the Moon-forming event, whose timing is still debated -- see the discussion in \citet{Thiemens2019,Kruijer2021,Thiemens2021}. \\

The simultaneous growth of Mars and Jupiter implies that TPF should no longer be studied in isolation, and that the growth of the gas giants should be included in dynamical models, or at least considered. The growing gas giants can have a profound effect on the composition of the inner Solar System as they scatter nearby planetesimals into the terrestrial region where gas drag circularises their orbits \citep{RaymondIzidoro2017}. The recent analyses of samples from asteroid (162173) Ryugu returned by JAXA's Hayabusa2 mission suggest that Ryugu and CI chondrites formed in the same region of the protoplanetary disc (REF), probably far away from the Sun. This has led \citet{Nesvorny2024} to perform a similar study to that of \citet{RaymondIzidoro2017} wherein they numerically quantify the implantation amount of CI-like material in the Uranus-Neptune region of the protoplanetary disc to the main asteroid belt. Potential further tentative evidence for the implantation scenario comes from purported early volatile accretion onto the angrite parent body \citep{Sarafian2014,Sarafian2017} \citep[c.f.][]{RS2024}, while the composition of the asteroid belt implies either mixing of materials \citep{Walsh2011}, or at least some compositional gradient \citep{DeMeoCarry2014}. \\

The main motivation for this study is cosmochemical: to build a model of TPF that accounts for the compositional diversity of Earth, Mars and Vesta. The implantation model of \citet{RaymondIzidoro2017} has potential to explain these features. The author opines that this scenario can explain the small amount of {\it jovian} material accreted by the terrestrial planets rather than relying on pebble accretion \citep[c.f.][]{Schiller2018,Schiller2020}, and further account for the isotopic differences between the Earth and Mars, as well as explain the composition of (162173) Ryugu \citep{Nesvorny2024}. A further motivation is provided by the idea that the CO, CV and CM carbonaceous chondrites could have formed in the main asteroid belt rather than beyond Jupiter \citep{Marrocchi2018}, for which dynamical models would predict a higher incorporation efficiency than if these formed beyond Jupiter. Here I take the work of \citet{RaymondIzidoro2017} to the next step. Not only do I trace how outer Solar System material finds its way into the inner Solar System, and what the implantation efficiency is, but I also track how (most of) this material is incorporated into the growing terrestrial planets. Recent advances in dynamically modelling the growth of the giant planets have placed some model constraints on the growth times of Jupiter and Saturn when these formed together \citep{Lau2024,Raorane2024}, so that the time has come for an in-depth study of how giant planet formation affects TPF. Regrettably, \citet{Raorane2024} had difficulty forming ice giant analogues, so that, unlike \citet{Nesvorny2024}, at present I will not study the implantation of distant ($>10$~au) material into the main asteroid belt.\\

The argument can be made that the dichotomy came to exist before the formation of Jupiter \citep{BM2020} and that the physical separation must have occurred at or beyond where Jupiter is today. However, if it is indeed possible that the CO, CV and CM carbonaceous chondrites could have formed in the main asteroid belt rather than beyond Jupiter then this warrants exploration. As such, in this study I will also explore the effect of having the boundary between {\it terrestrial} and {\it jovian} group material be closer to the Sun than 5~au.\\

This paper is part of a series of manuscripts detailing the implantation process, and how much {\it jovian} material ends up incorporated into the terrestrial planets and the main asteroid belt. Here I focus on the first 5 Myr of evolution, when the protosolar disc was still around. Future publications will reveal the long-term evolution of the system.

\section{Methods}
\begin{figure}
\resizebox{\hsize}{!}{\includegraphics{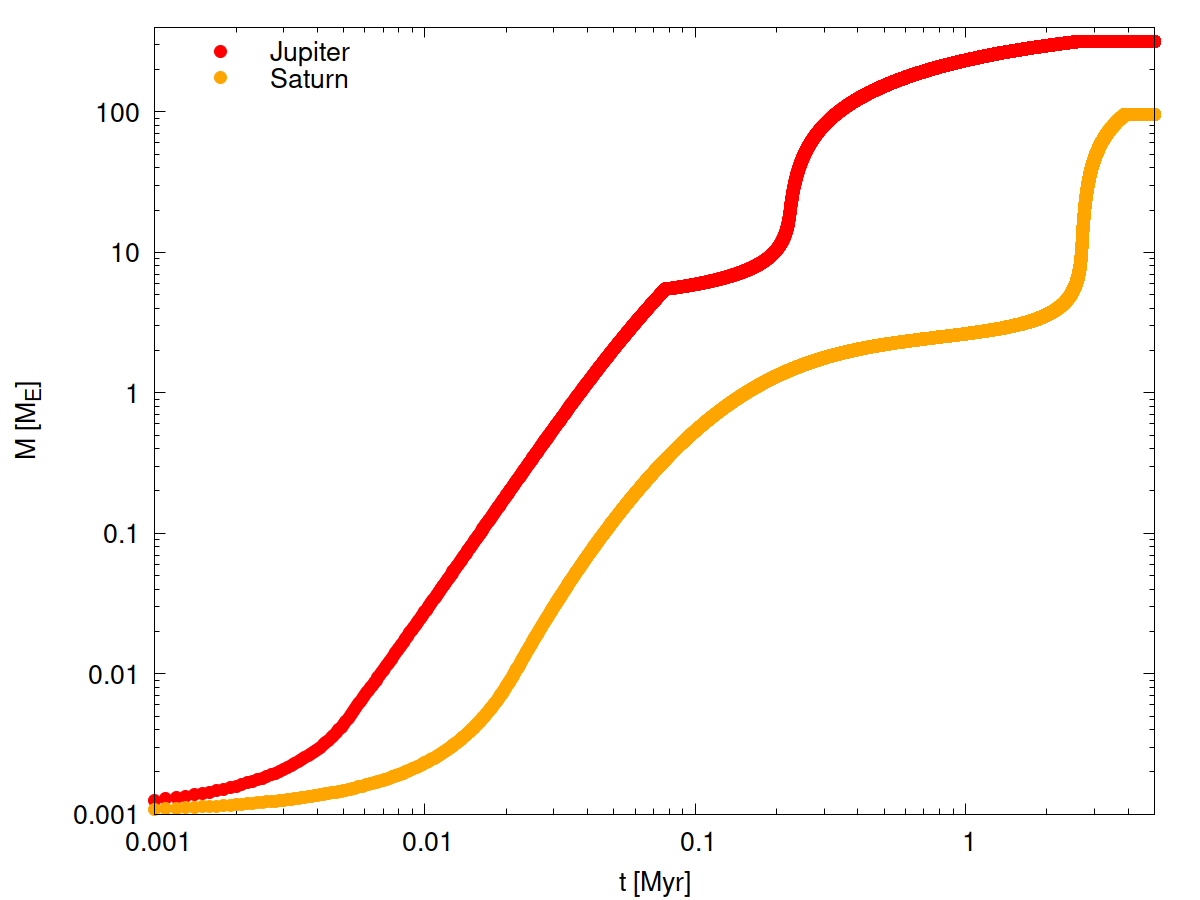}}
\caption{Growth of the masses of Jupiter and Saturn as a function of time adopted in this study.}
\label{fig:jsg}%
\end{figure}
\subsection{Growth of the gas giants}
The first thing my model requires is to have a presciption for the growth of the giant planets. \citet{RaymondIzidoro2017} began Jupiter and Saturn as planetary cores with a mass of 3~Earth masses ($M_\oplus$) each, and then grew their masses to their current values on a timescale of 0.1 million years, although faster and slower growth timescales were also tested. Here I want to have the growth timescales of Jupiter and Saturn match that of dynamical simulations of pebble accretion, but also not be too different from that adopted by \citet{RaymondIzidoro2017} for easier comparison. \\

Following \citet{BM2020} and guided by the extensive simulations of giant planet formation in the solar system with pebble accretion by \citet{Lau2024,Raorane2024} I generated growth tracks of the giant planets using the N-body integrator SyMBA \citep{Duncan1998}, which was modified to include the effects of pebble accretion, gas accretion and damping forces from the gas disc \citep{Matsumura2021}. These highly simplified simulations started with two Ceres-mass planetesimals at 5.6 au and 7.5 au which were subjected to a flux of pebbles; I could have chosen larger initial masses, but I found that this initial mass gave growth tracks that satisfied the growth rates from \citet{RaymondIzidoro2017} and \citet{Raorane2024} because the initial growth is slow, causing a slight delay until there is gas envelope accretion of the Jupiter analogue. The disc parameters, pebble flux and temporal disc evolution were the same as those in \citet{Raorane2024}: the initial mass of the protoplanetary disc was 0.05 solar masses ($M_\odot$), the diffusion timescale of the gas was set to 0.5~million years and the slope of the temperature gradient in this region of the disc was that of the irradiated regime, i.e. $|d\ln T/d\ln r| = 3/7$ \citep{Ida2016}. The temporal mass accretion onto the star and the pebble mass flux are given by
\begin{eqnarray}
\dot{M}_* &=& \frac{7}{13}\left(\frac{M_d}{M_\odot}\right)\left(\frac{1 {\rm yr}}{t_{\rm diff}}\right)\left(\frac{t}{t_{\rm diff}}+1\right)^{-20/13}\,M_\odot\,{\rm yr}^{-1}\nonumber \\
\dot{M}_F &=& 1.51\times 10^{-3} \left(\frac{10^{-3}}{\alpha_{\rm acc}}\right) \left(\frac{\dot{M}_{*}}{10^{-8}\,M_\odot\,{\rm yr}^{-1}}\right) \left(\frac{t}{t_{\rm pff}}\right)^{-8/21}\, M_{\oplus}\, {\rm yr}^{-1},
\end{eqnarray}
where I have substituted nominal disc and pebble radius parameters \citep{Matsumura2021,Lau2024,Raorane2024}. Here $\alpha_{\rm acc}$ is the accretion viscosity and $t_{\rm pff}$ is the pebble formation time \citep{Ida2016}. The pebble accretion prescription and disc model of \citet{Ida2016} was used assuming perfect accretion efficiency. \\

Gas envelope accretion was computed as described in \citet{Raorane2024}, where I used the same input parameters as that study. In short, the envelope mass evolves as \citep{Ida2018}

\begin{equation}
    \frac{dM_{\rm env}}{dt} = {\rm min}\Bigl[\frac{M_{\rm core}}{\tau_g},\dot{M}_*,f_{\rm gap}\dot{M}_* \Bigr],
\end{equation}
where $f_{\rm gap}$ is a reduction factor due to gap opening of the growing planet \citep{Ida2018}, and is given by

\begin{equation}
f_{\rm gap} = \frac{0.031}{(1+0.04K)h^4\alpha_{\rm acc}}\Bigl(\frac{m_p}{M_*}\Bigr)^{4/3}.
\end{equation}
where $m_p$ is the mass of the planet and where $\tau_g$ is the gas accretion timescale, given by

\begin{equation}
    \tau_g = 10^8 \Bigl(\frac{M_\oplus}{M_{\rm core}}\Bigr)^3  \Bigl(\frac{\kappa_{\rm gr}}{1\, {\rm cm}^2\,{\rm g}^{-1}}\Bigr)\: {\rm yr}
\label{eq:gasenv}
\end{equation}
where $M_{\rm core}$ is the core mass and $\kappa_{\rm gr}$ is the grain opacity \citep{Ikoma2000}; for this study I used an opacity of 1~cm$^{2}$~g$^{-1}$. The factor $K$ is related to gap opening and is given by \citep{Kanagawa2018} 

\begin{equation}
K=\Bigl(\frac{m_p}{M_*}\Bigr)^2h^{-5}\alpha_{\rm turb}^{-1}, 
\end{equation}
where $\alpha_{\rm turb}$ represents the strength of the local disc turbulence, which I set to $\alpha_{\rm turb}=10^{-4}$. Gap opening becomes important when $K = 25$. \\

The time step in the simulation was 0.1~yr and the sims were run for 5 Myr. I switched off the effect of planetary migration to mimic a purported disc pressure maximum at 5~au \citep{BM2020} and to not have the growing gas giants encounter each other, or to migrate all the way to terrestrial planet region. The outcome of this setup seems inconsistent with the results from \citet{Raorane2024}, wherein massive gas giants frequently experienced mutual encounters and Jupiter analogues were on average farther away than Saturn analogues, but the purpose of this study is to establish the implanetation efficiency and how {\it jovian} material is incorporated into the terrestrial planets.\\

The growth time of Jupiter calculated here is similar to what was used by \citet{RaymondIzidoro2017}, and although it is fast it is still consistent with that reported in \citet{Raorane2024}. A follow-up study using a slower growth of Jupiter is in progress. \\

The mass evolution of both planets is shown in Fig.~\ref{fig:jsg}. Both planets grow from pebble accretion until approximately 6~$M_\oplus$, after which gas accretion sets in \citep{Matsumura2021}. The planet's masses were truncated to their current values. The growth rate of Jupiter once it reaches 3~$M_\oplus$ to its current value is similar to that employed by \citet{RaymondIzidoro2017}, while in my simulation Saturn's growth is much slower. The growth rate of Jupiter is much faster but still statistically consistent with that found in \citet{Raorane2024}, while the growth rate of Saturn is mostly consistent with that study. Although the growth rate of Jupiter is fast, the initial simulation time is somewhat arbitrary when compared to the absolute age of the Solar System and the meteorite record, with a probable window comparable to the growth of the non-carbonaceous iron meteorites, i.e. $\sim$0.5~million years \citep{Kruijer2014}. The effect of a slower growth of Jupiter will be left for future work. From these growth tracks I extracted orbital elements, masses and radii of the gas giants. Both gas giants became slightly eccentric due to their mass growth changing the centre of mass of the system, with their mutual perturbations also inducing some finite value of eccentricity. \\

\subsection{GPU simulations of TPF with growing gas giants}
Next, I ran GPU N-body simulations using GENGA of a planetesimal disc in the presence of gas from the protplanetary disc that was subjected to the growing Jupiter and Saturn. At present GENGA does not have pebble accretion implemented so that the gravitational effect of the growing gas giants on the inner Solar System has to be approximated. GENGA can read and interpolate masses, radii and orbital elements of bodies from a table, so I used that here; I read the masses, radii, semi-major axes, eccentricities, and argument of perihelion of the gas giants every 100~yr. Their mutual inclinations are zero, and their nodes were fixed. With the amount of High Performance Computing (HPC) time I had available I was able to run seven sets of eight simulations of the evolution of a swarm of planetesimals subjected to the growing gas giants. The simulation sets differed in their initial conditions in total planetesimal mass and planetesimal number. In Table~\ref{tab:init} I list the different initial conditions of the sets of simulations. Each individual simulation started with 16k to 61k self-gravitating planetesimals situated between 0.5~au and 8.5~au, and two Ceres-sized bodies at 5.6 au and 7.5 au that are artificially grown to become Jupiter and Saturn. Planetesimals usually had a radius between 300 and 600~km and are fully interacting, they were uniformly randomly distributed in semi-major axis in each annulus (inner and outer), had uniformly random eccentricities $\leq 0.001$ and inclinations $\leq 0.5^\circ$, and uniformly randomised other angles. The mass of the planetesimal disc that is a proxy for {\it terrestrial} material is listed as $m_{\rm inner}$ and typically consists of 8k to 10k planetesimals; similarly the mass of the outer planetesimal disc located up to 8.5~au is denoted $m_{\rm outer}$ and consists of 8k to 51k planetesimals. The distance at which I assumed the composition of the bodies to change from {\it terrestrial} type to {\it jovian} type is listed as 'Change'; \citet{RaymondIzidoro2017} initially assumed a change in composition occurred at 4~au. This is a sharp transition between the two groups of material, but this approach makes tracking the evolution of material in each group much easier than when the gradient is more gradual. If there is an empty region, to mimic an initially empty asteroid belt \citep{RI2017}, the inner and outer edges are specified in 'Gap'. The initial number of planetesimals and number of simulations were limited by the amount of HPC hours I was able to obtain.\\

Simulations were run for 5 Myr with a time step of 0.01 yr; the remaining evolution will be discussed in a forthcoming publication. The surface density of the protoplanetary disc declined exponentially with an e-folding time of 1~Myr; its initial value was 1700~g~cm$^{-2}$ at 1 au with an inner edge of 0.4~au, and the surface density declines with distance as $r^{-1}$. No migration was imposed in order to avoid a pileup of embryos in the Mercury region \citep{Woo2021}, and because there was a pressure maximum near Venus that would impede this migration \citep{Woo2023}. GENGA has the option to only apply the gas forces and torques to bodies less massive than a threshold mass, dubbed $m_{\rm Giant}$. In my simulations most of the planetesimals have an initial mass comparable to the seeds of Jupiter and Saturn. The latter two bodies must not be subject to the gas forces because it causes errors in the interpolation of their growth tracks, hence $m_{\rm Giant}$ must be smaller than their initial masses. To model this accurately I would have to start and stop the simulations and adjust $m_{\rm Giant}$ at regular intervals, which causes strange behaviour in the migration. As such, I opted for simplicity and have none of the bodies be subjected to migration. Planetesimals with mass $m<m_{\rm Giant}$ experienced gas drag and eccentricity and inclination damping from the gas disc as implemented in GENGA \citep{grimmetal2022}; details of their prescriptions are listed in \citet{Woo2021}. The avoidance of a pileup of embryos in the Mercury region is addressed by the works of \citep{Clement2021a,Clement2021b,Clement2021c} and in \citet{Woo2023}. In future studies I will address this problem in a similar manner. In the GPU simulations bodies were removed closer than 0.3 au or farther than 300 au from the Sun, or when there was a collision. Perfect mergers are assumed whenever there was a collision between bodies.

\begin{table}
\begin{tabular}{r|ccccc}
Set & $m_{\rm inner}$ & $m_{\rm outer}$ & Change & Gap (range) & Bodies \\ \hline \\
334 & 3 & 3 &4 & - & 16k \\
232 & 2 & 3 & 2 &- & 61k \\
233 & 2 & 3 & 3& - & 47k \\
234 & 2 & 3 & 4& - & 61k \\
235 & 2 & 3& 5&  - & 47k \\
232g4 & 2 & 3 & 4 & 2-4 & 41k \\
232g5 & 2 & 3 & 5 & 2-5 & 30k
\end{tabular}
\caption{Simulation parameters. The masses are total planetesimal masses in Earth masses. The column 'Change' implies at what distance (in au) I consider the composition of the planetesimals to be predominantly inner versus outer solar system. The fourth column indicates whether there is an empty region, and where (in au). The last column is self-explanatory.}
\label{tab:init}
\end{table}

\section{Results}
In this work I focus on how the implantation of planetesimals from the {\it jovian} group affects the bulk composition of the terrestrial planets, beginning with planetary embryos here and going all the way to the final terrestrial planets in a subsequent publication. As such, I will not describe the detailed dynamics of the swarm of planetesimals because it is very similar to that of \cite{Woo2021} as well as that of \citet{RaymondIzidoro2017}. I refer the reader to those works for an in-depth exploration of the dynamics, but I will highlight specific aspects.
\begin{figure}[]
\resizebox{\hsize}{!}{\includegraphics{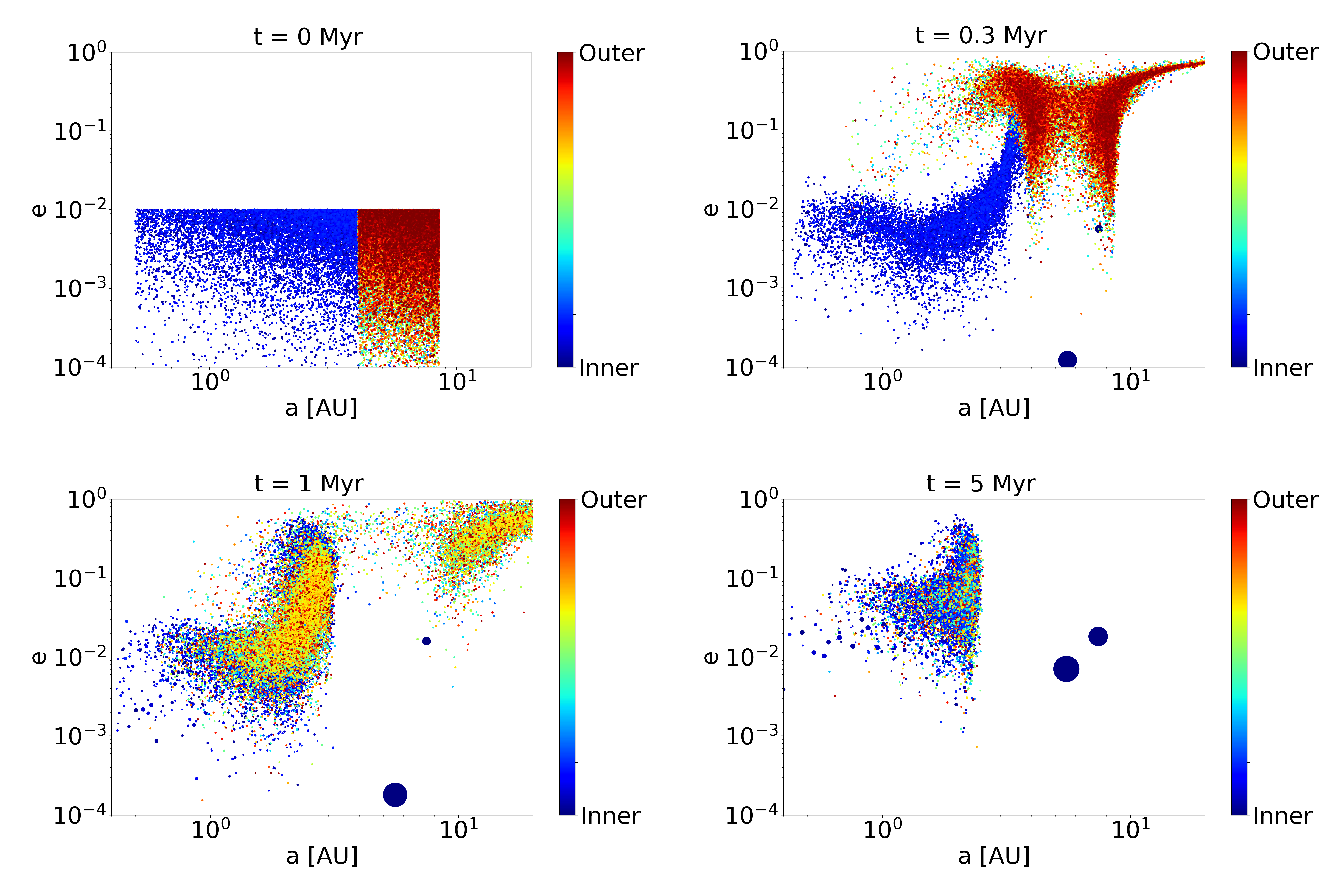}}
\caption{Snapshots of the evolution of planetesimal implantation from a simulation from set 234 starting with 61k planetesimals as the gas giants grow. Terrestrial material is colour coded green-blue and jovian material as yellow-brown. After 0.3 million years the growing Jupiter is scattering planetesimals away, some of which are circularised by gas drag into the inner solar system. After 1 Myr the inner solar system is mixed, and by 5 Myr some embryos can be seen.}
\label{fig:implant}%
\end{figure}

\subsection{General evolution and planetesimal implantation}
\begin{figure}
\includegraphics[height=0.18\textheight,width=0.5\textwidth]{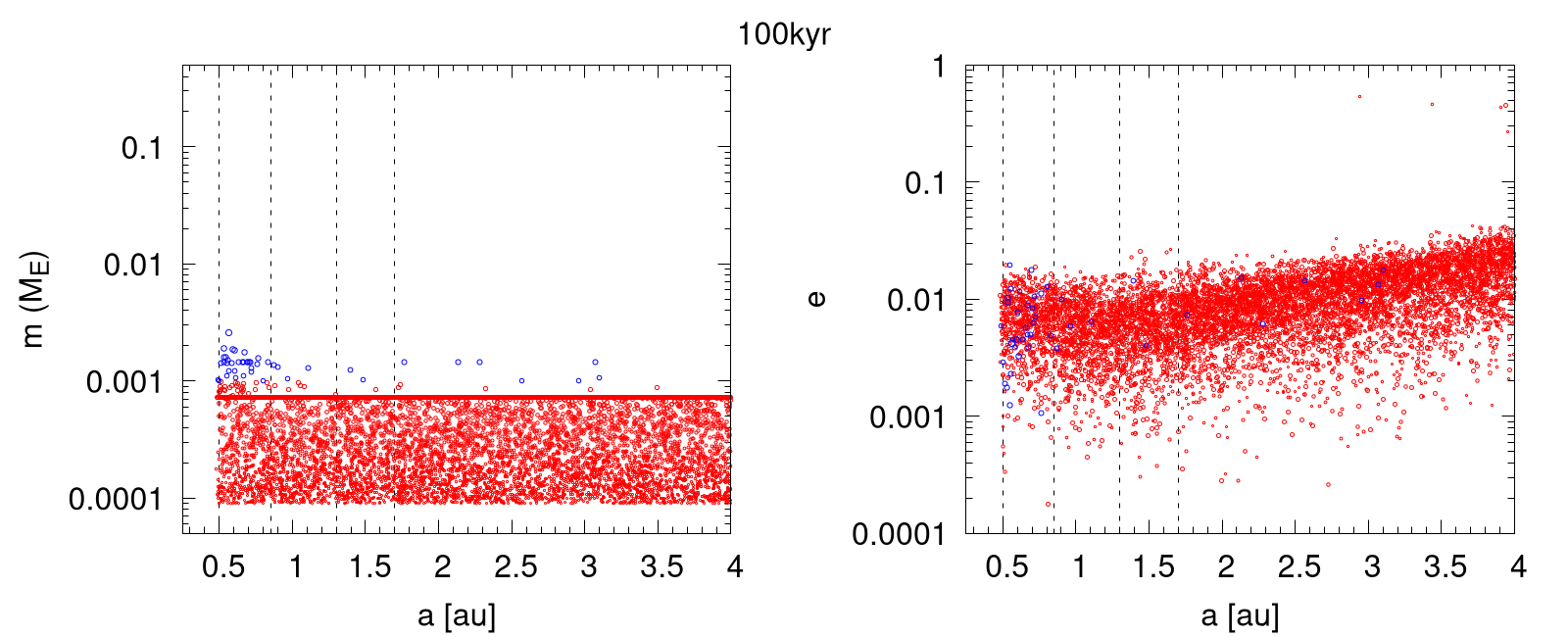}\\
\includegraphics[height=0.18\textheight,width=0.5\textwidth]{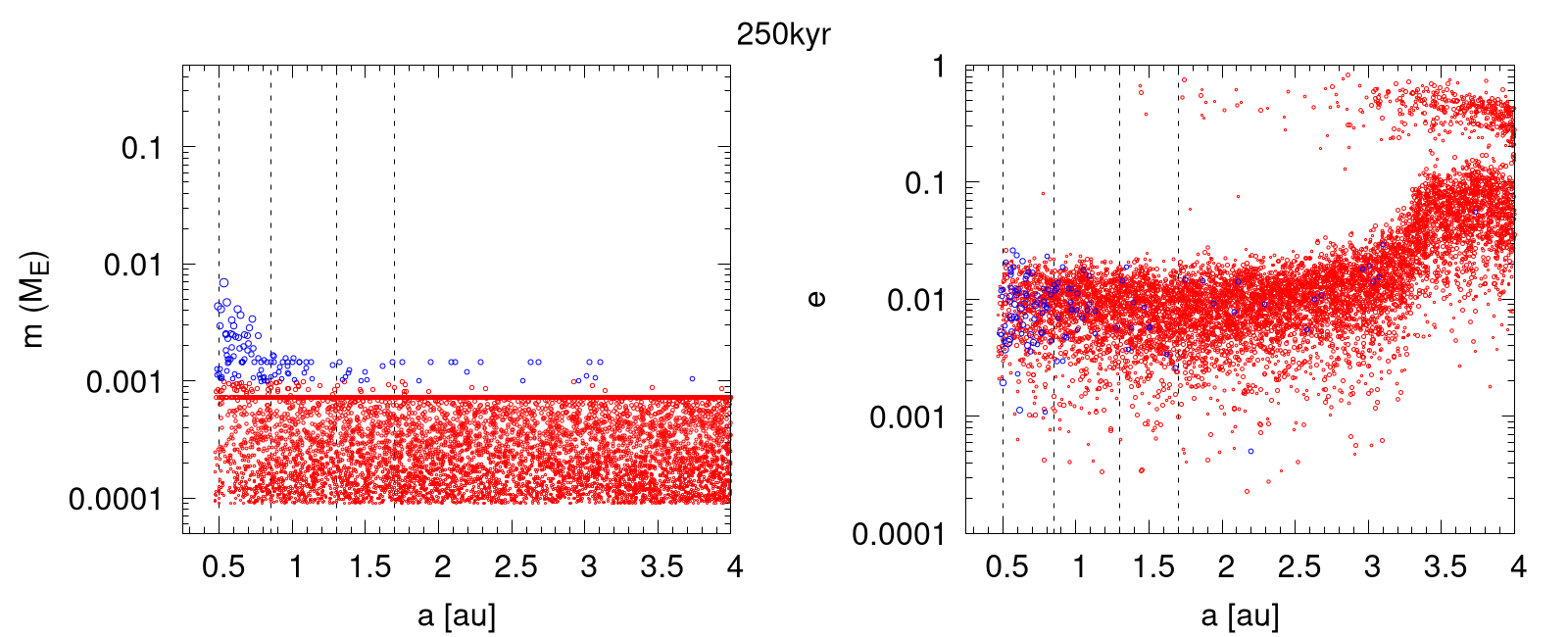}\\
\includegraphics[height=0.18\textheight,width=0.5\textwidth]{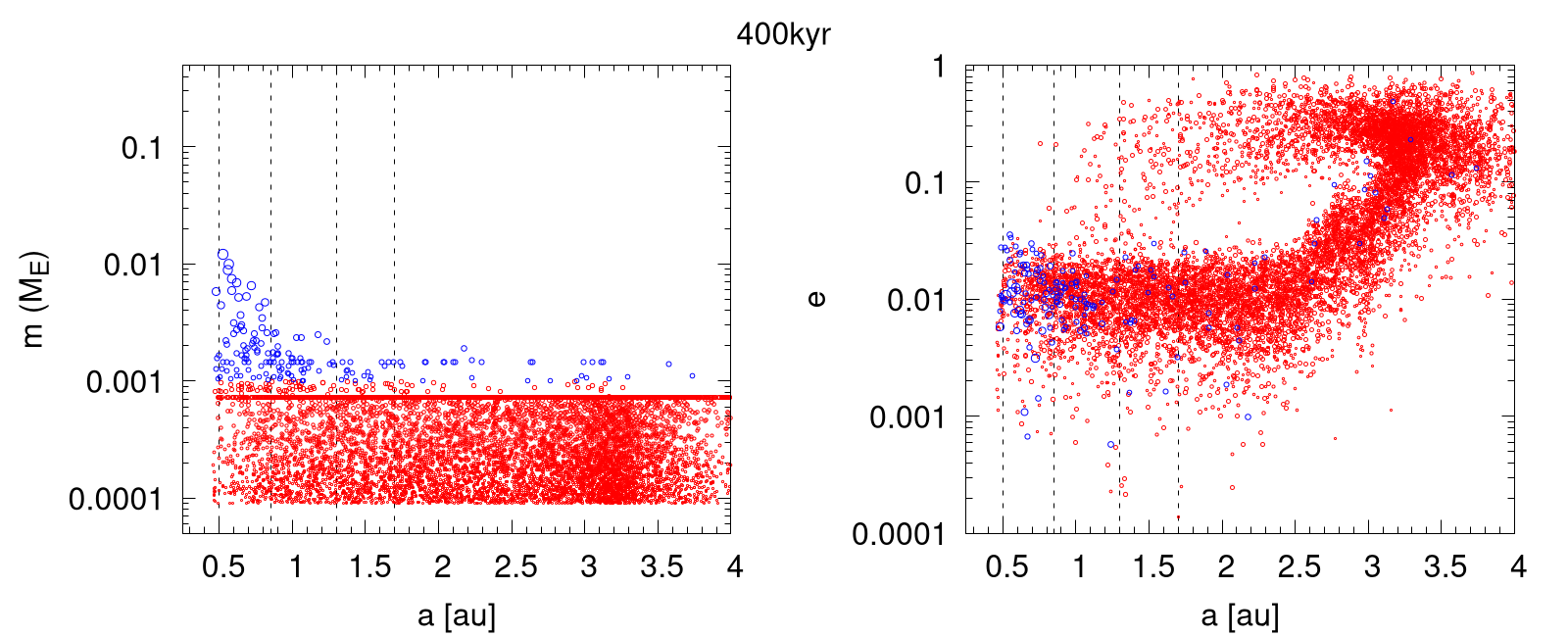}\\
\includegraphics[height=0.18\textheight,width=0.5\textwidth]{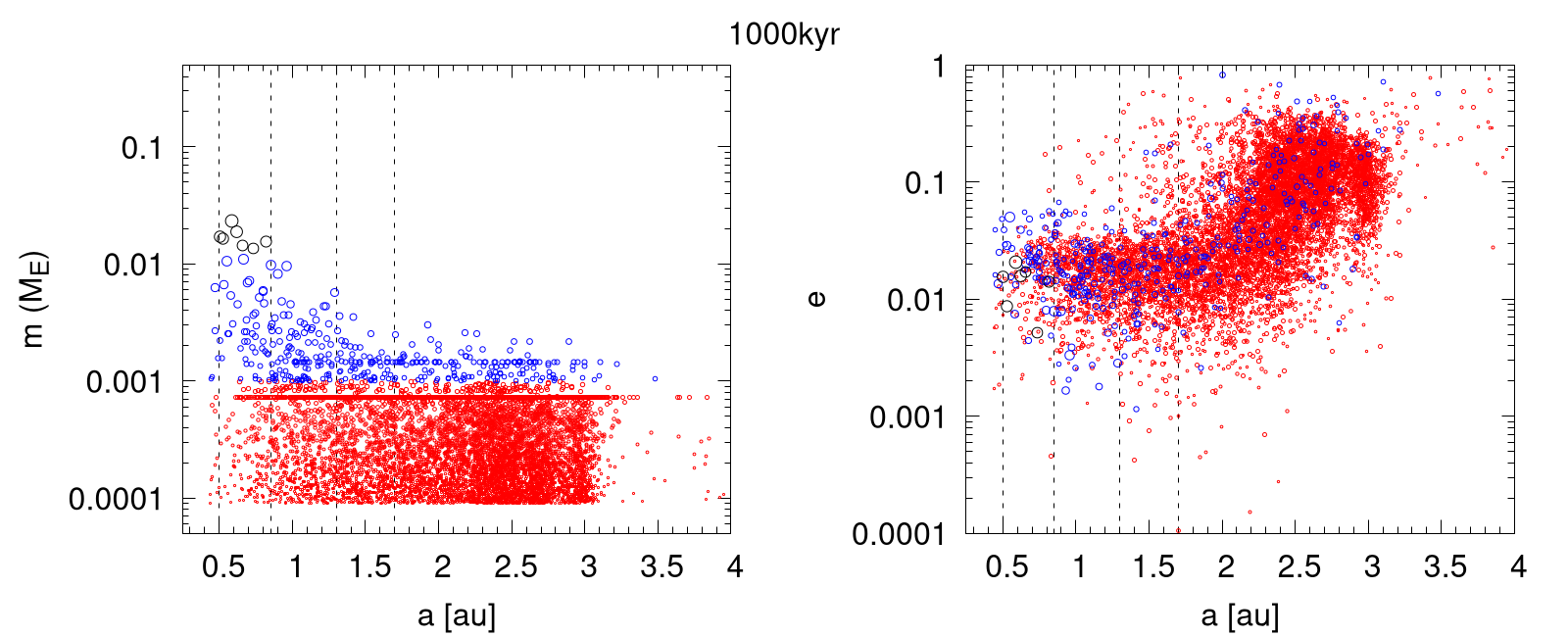}\\
\includegraphics[height=0.18\textheight,width=0.5\textwidth]{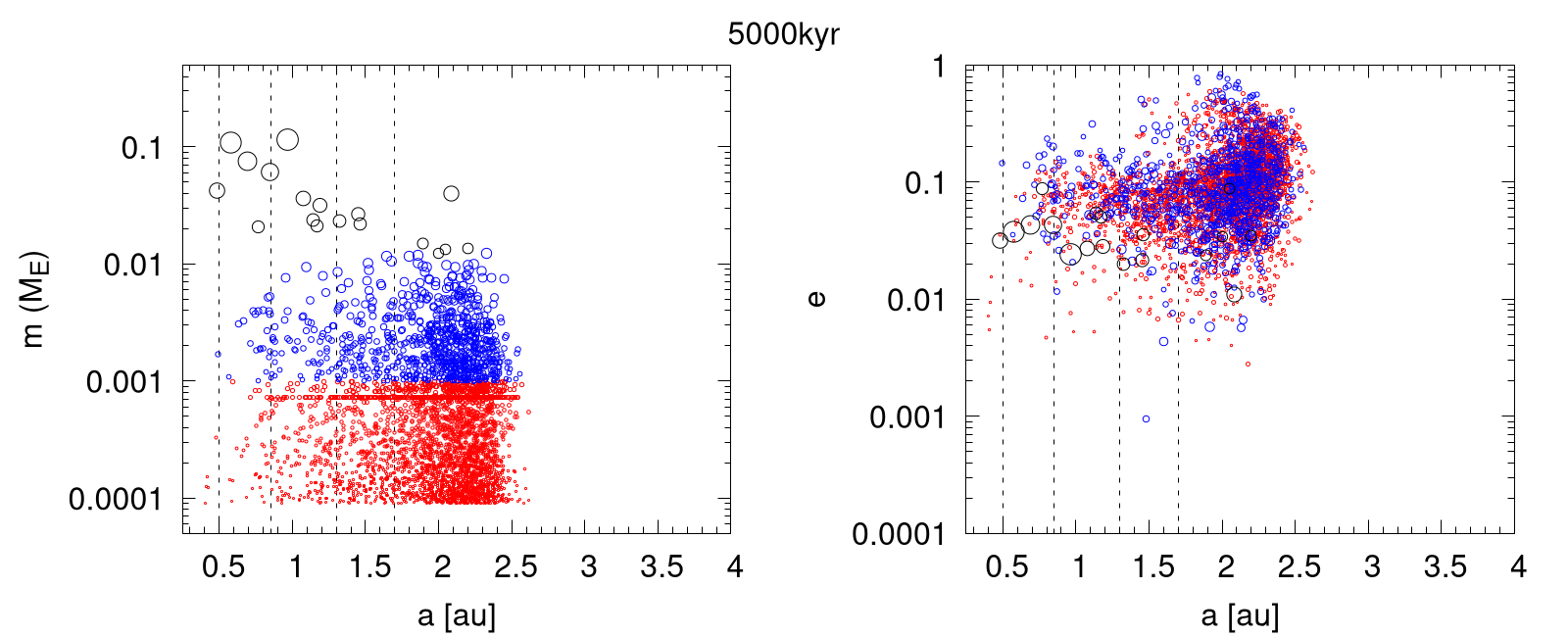}
\caption{Snaphots of the evolution of a simulation from set 232. Implantation is ongoing at 250~kyr, while after 400~kyr the $\nu_5$ secular resonance begins to sweep across the asteroid belt. Red dots: bodies with mass $<10^{-3}$~$M_\oplus$, blue for $10^{-3}<m<10^{-2}$~$M_\oplus$, and black for more massive bodies.}
\label{fig:avsme}%
\end{figure}
Snapshots of a single simulation from the first simulation set is displayed in Fig.~\ref{fig:implant}; the simulation began with approximately 16,000 planetesimals plus the seed embryos for the gas giants. The snapshots show the semi-major axis and eccentricitry of planetesimals (blue-green and brown dots), and the growing gas giants (large blue dots). The colour coding of the planetesimals distinguishes {\it terrestrial} group (blue-green) and {\it jovian} group (yellow-brown) material; in this simulation the assumed initial transition between both reservoirs is at 4~au. As Jupiter grows it begins to scatter the planetesimals in its vicinity, keeping their aphelion or perihelion pegged to the planet, which creates the 'wings' seen in the top-right and bottom-left panels. In the top-right panel, after 0.3 million years of evolution, there are brown dots with semi-major axis $a\sim2$~au and $e > 0.3$. These bodies have been dynamically decoupled from Jupiter due to gas drag and they will be implanted into the asteroid belt or the terrestrial planet region. After 1~Myr of evolution (bottom-left panel) the inner Solar System has become a mixture of green-blue and yellow-brown, with small planetary embryos beginning to form near 0.5~au. After 5~Myr (bottom-right panel) the asteroid belt beyond 2.5~au has been mostly emptied out, the planetesimals with $a>5$~au that were being scattered by the gas giants have also gone, the remaining swarm of planetesimals in the inner Solar System has a mixed composition, and planetary embryos are clearly visible in the region between 0.5~au and 1~au. The prediction from this simulation is that the material in the inner Solar System is thoroughly mixed.\\

A more detailed evolution of the same simulation is given in Fig.~\ref{fig:avsme}, which are five two-panel plots at different times. The left column shows the semi-major axis versus mass, the right column the semi-major axis versus eccentricity. The symbol size is proportional to a body's radius. Red dots are for bodies with a mass $<10^{-3}$~$M_\oplus$, black for masses $>0.01$~$M_\oplus$, and blue for bodies whose masses fall in between these values. In the second row, after 0.25 million years, there are planetesimals with very high eccentricities $e>0.3$ and semi-major axes $a>2$~au. These are again the {\it jovian} group planetesimals that are being scattered inwards by the growing Jupiter. In the next panel, after 0.4 million years, the outer part of the asteroid belt with $a>3$~au has rising eccentricities, and the outer region is mostly empty after 1 million years of evolution in the fourth row. In the last row, after 5~million years of evolution, the region beyond 2.5~au is empty. The reason for this emptying out is explained in \cite{Woo2021}: it is caused by the sweeping of the $\nu_5$ secular resonance across the asteroid belt as the gas disc's mass decreases and the precession rate of the asteroids decreases. Those asteroids for which $\dot{\varpi} -g_5 \approx 0$, where $g_5$ is the secular eigenfrequency corresponding to Jupiter's perihelion precession, encounter the secular resonance and have their eccentricities pumped to high values \citep{Ward1981}, so that their perihelia decrease and the higher gas surface density closer to the Sun circularises their orbits in the terrestrial planet region. By the end of the simulation several Mars-mass objects have formed in the Venus region, with many Ceres to lunar-mass objects remaining.\\

\begin{figure}
\resizebox{\hsize}{!}{\includegraphics{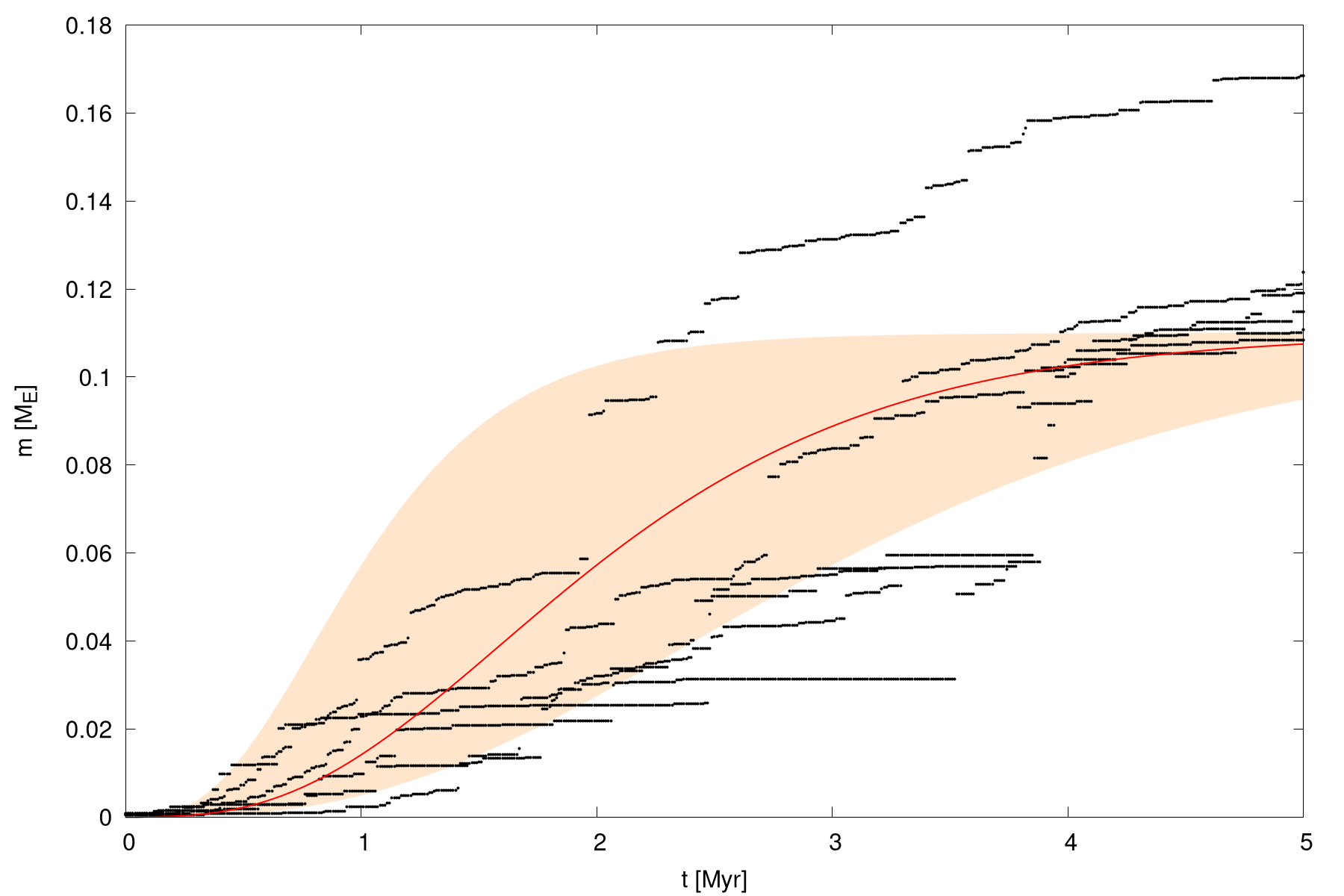}}
\caption{Growth tracks of embryos from the 232 and 334 sets that grow to Mars' mass. Most of these are close to 1~au, so they are not actual Mars analogues. Near the current position of Mars embryo growth is generally too slow \citep{Woo2021}.}
\label{fig:mars}%
\end{figure}
\subsection{Embryo growth timescales}
The average time it takes to grow to a 0.01~$M_\oplus$ embryo depends on the initial conditions. For the 232 case, in which bodies grow the fastest, the time taken is $1.5^{+2.0}_{-1.3}$~million years (2$
\sigma$), which is comparable to what was reported in \citet{Woo2021} (in particular their Fig. 8). For the 235 case, in which growth is the slowest, the time taken is about $2.5^{+1.5}_{-2.4}$~million years. For the other sets of simulations the time taken to grow to a lunar mass falls in between. The growth timescale of Mars has been deduced from Hf-W chronology to be about 5~Myr \citep{DauphasPourmand2011}. In that work, the authors use the parametric growth expression of Mars as $M(t)=M_{\rm Mars}\tanh^3(t/\tau)$, with $\tau = 1.8 \pm 0.9$~million years. Following \citet{Woo2021} we have plotted this growth curve and several cases of planetary growth from sets 232 and 334 wherein a planetesimal collides with multiple bodies to become a Mars-sized embryo, or larger. As is clear, the growth rate of these embryos roughly spans the range of timescales advocated by \citet{DauphasPourmand2011}. However, none of these embryos resides near 1.5~au, where the growth timescale is generally too slow \citep{Woo2021}. At 5 million years the number of embryos per simulation more massive than 0.01~$M_\oplus$ is typically 15-30. 

\begin{figure*}
\resizebox{\hsize}{!}{\includegraphics{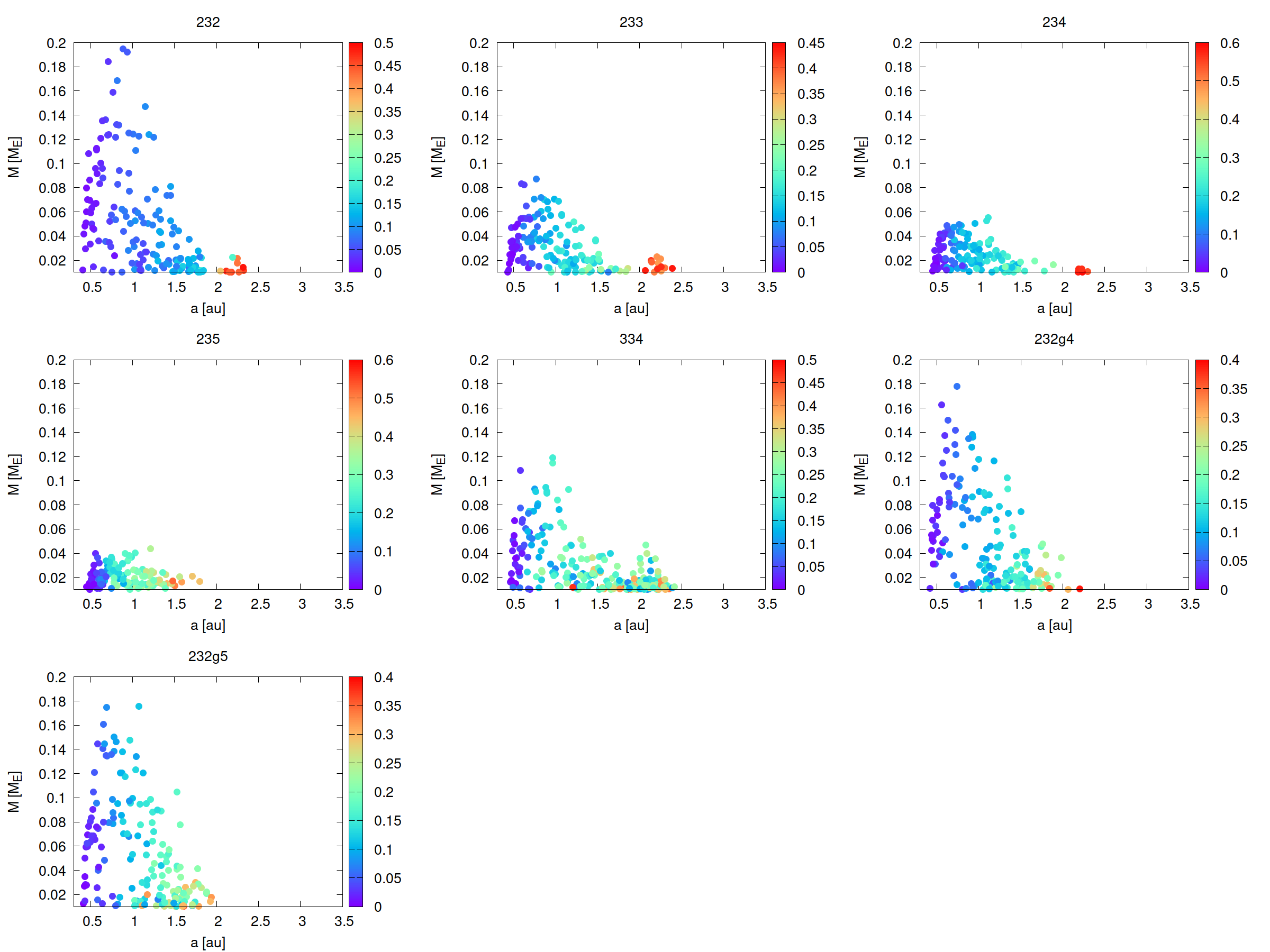}}
\caption{Embryo composition after 5 Myr. The panels show the final semi-major axis versus mass of the embryos for all seven sets of simulations. The colour coding depicts the fraction of embryo mass that comes from the {\it jovian} group. Note that there is a composition gradient, with the embryos close to 0.5~au having 5\% of their mass from the {\it jovian} region while this is closer to 20\% for embryos at 1.5~au.}
\label{fig:comp}%
\end{figure*}

\begin{figure}
\includegraphics[height=0.22\textheight,width=0.44\textwidth]{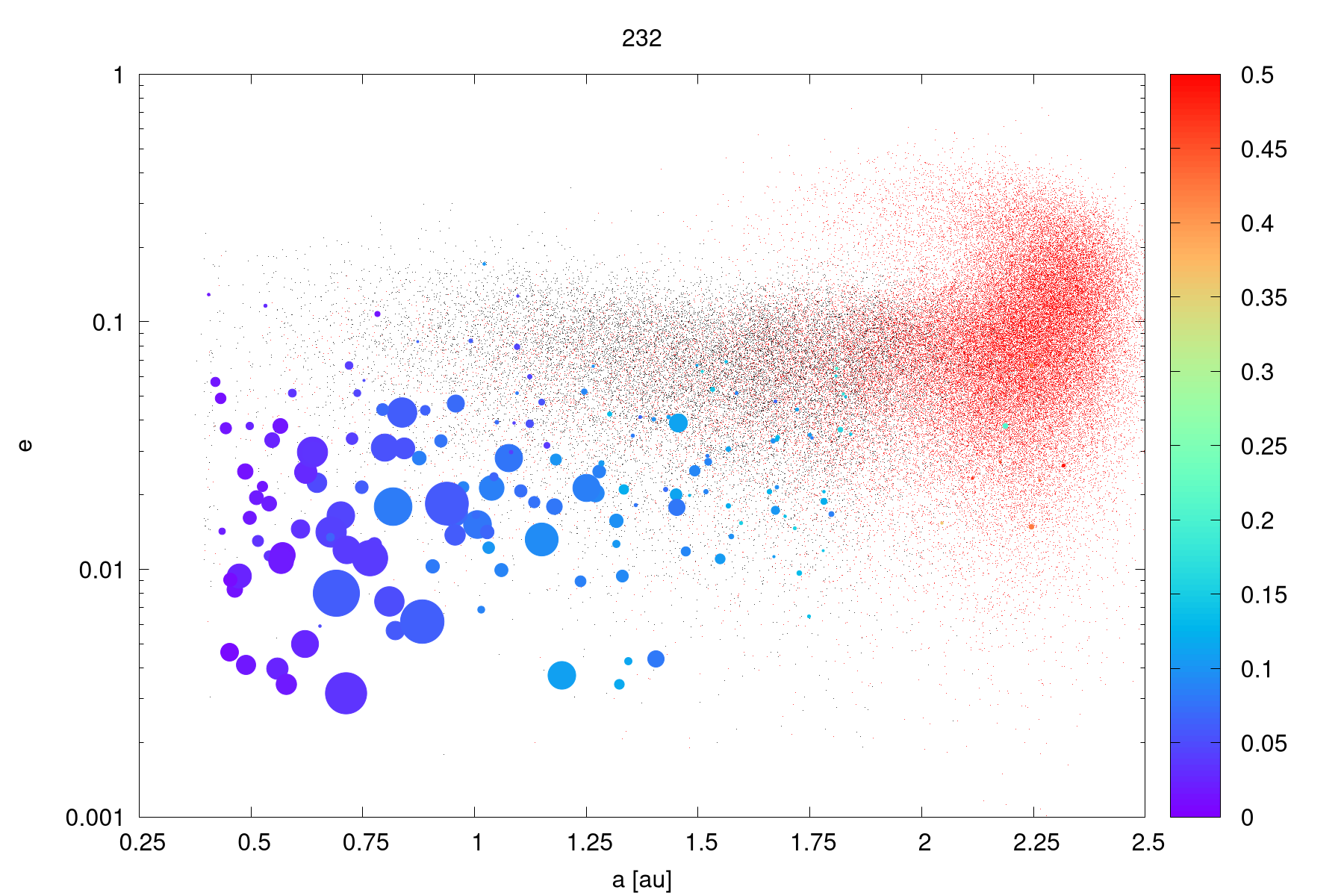}\\
\includegraphics[height=0.22\textheight,width=0.44\textwidth]{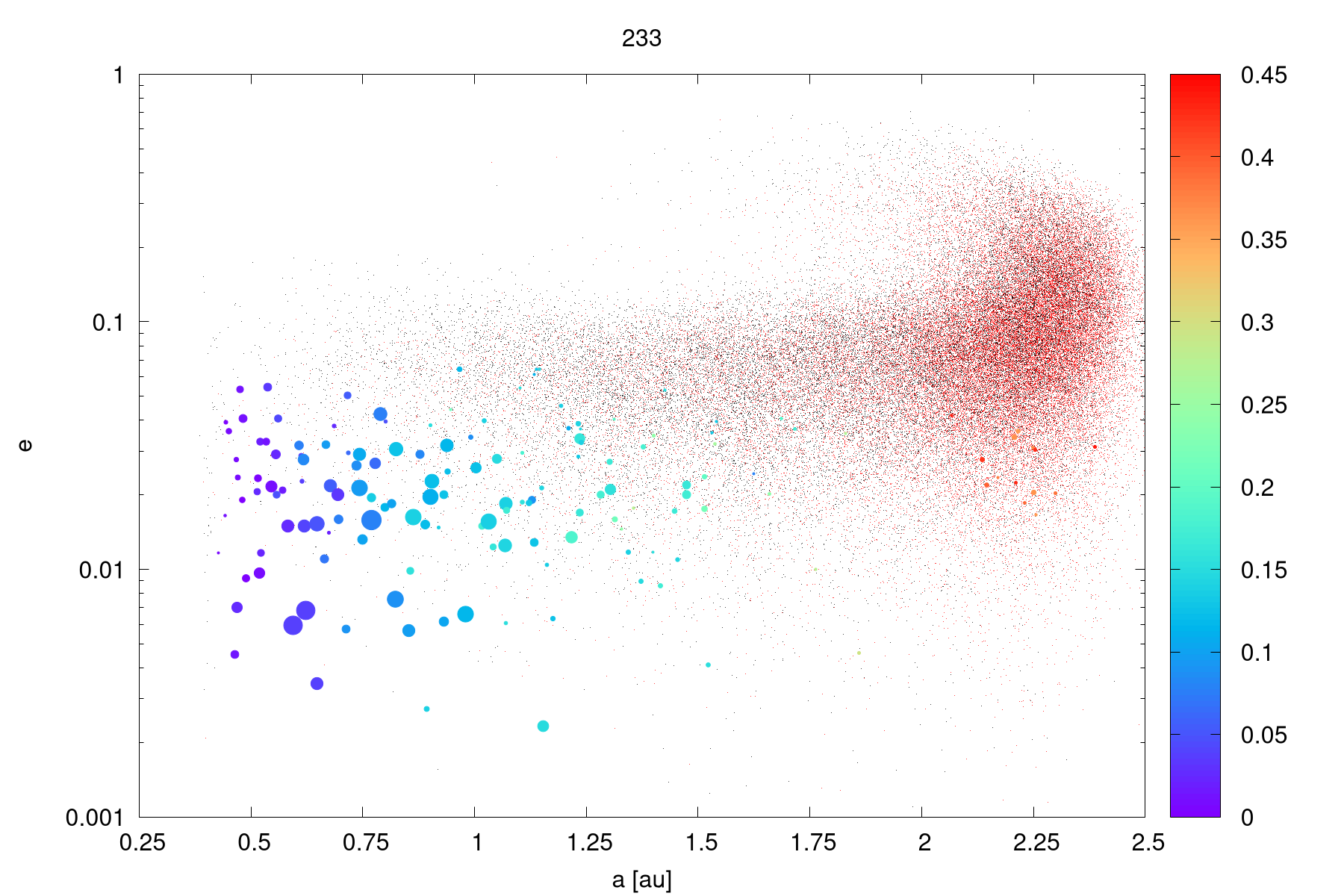}\\
\includegraphics[height=0.22\textheight,width=0.44\textwidth]{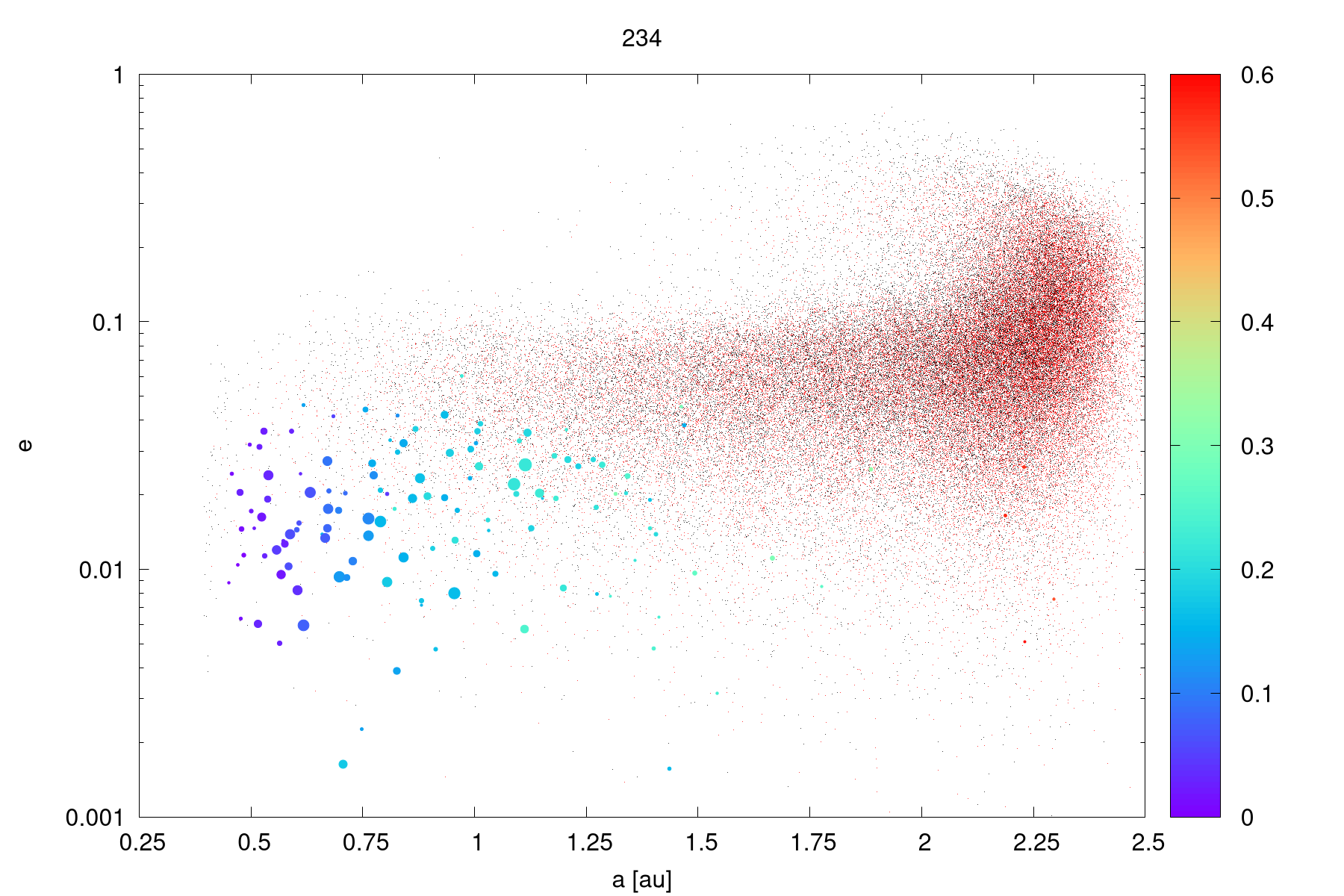}\\
\includegraphics[height=0.22\textheight,width=0.44\textwidth]{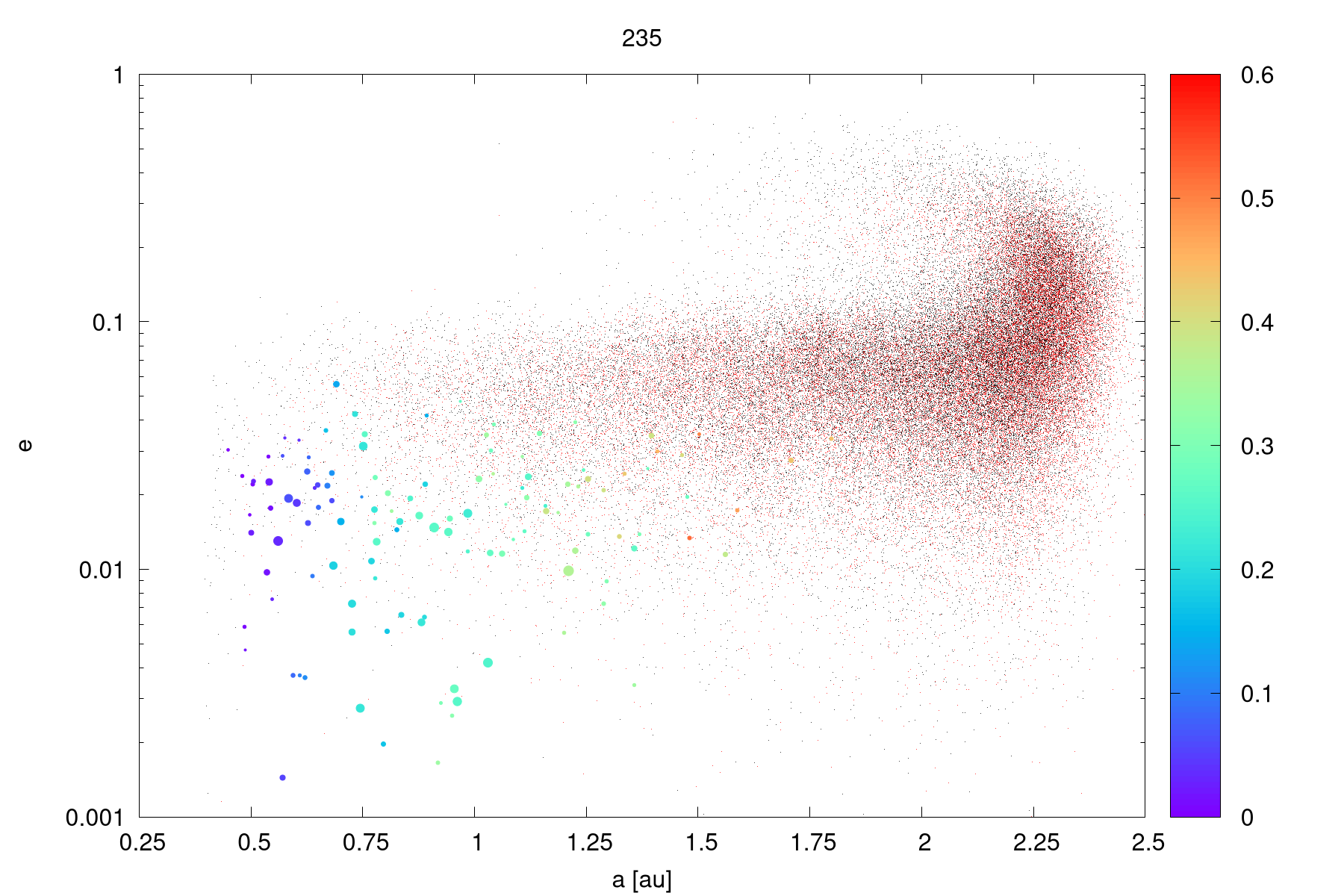}
\caption{Snapshots of the final semi-major axis and eccentricity distribution of embryos and planetesimals for the simulation sets 232, 233, 234 and 235, showing the decreasing efficency of embryo formation. Apart from the top panel, the planetesimal population closer than 2.5~au is completely mixed (black for {\it terrestrial} group material, red for {\it jovian} group material). Almost no planetesimals reside on orbits with $a>2.5$~au. The colourbar shows the fraction of {\it jovian} group material incorporated in the embryos.}
\label{fig:snapcomp}%
\end{figure}

\subsection{Embryo composition, feeding zones and implantation efficiency}
\begin{figure*}
\resizebox{\hsize}{!}{\includegraphics{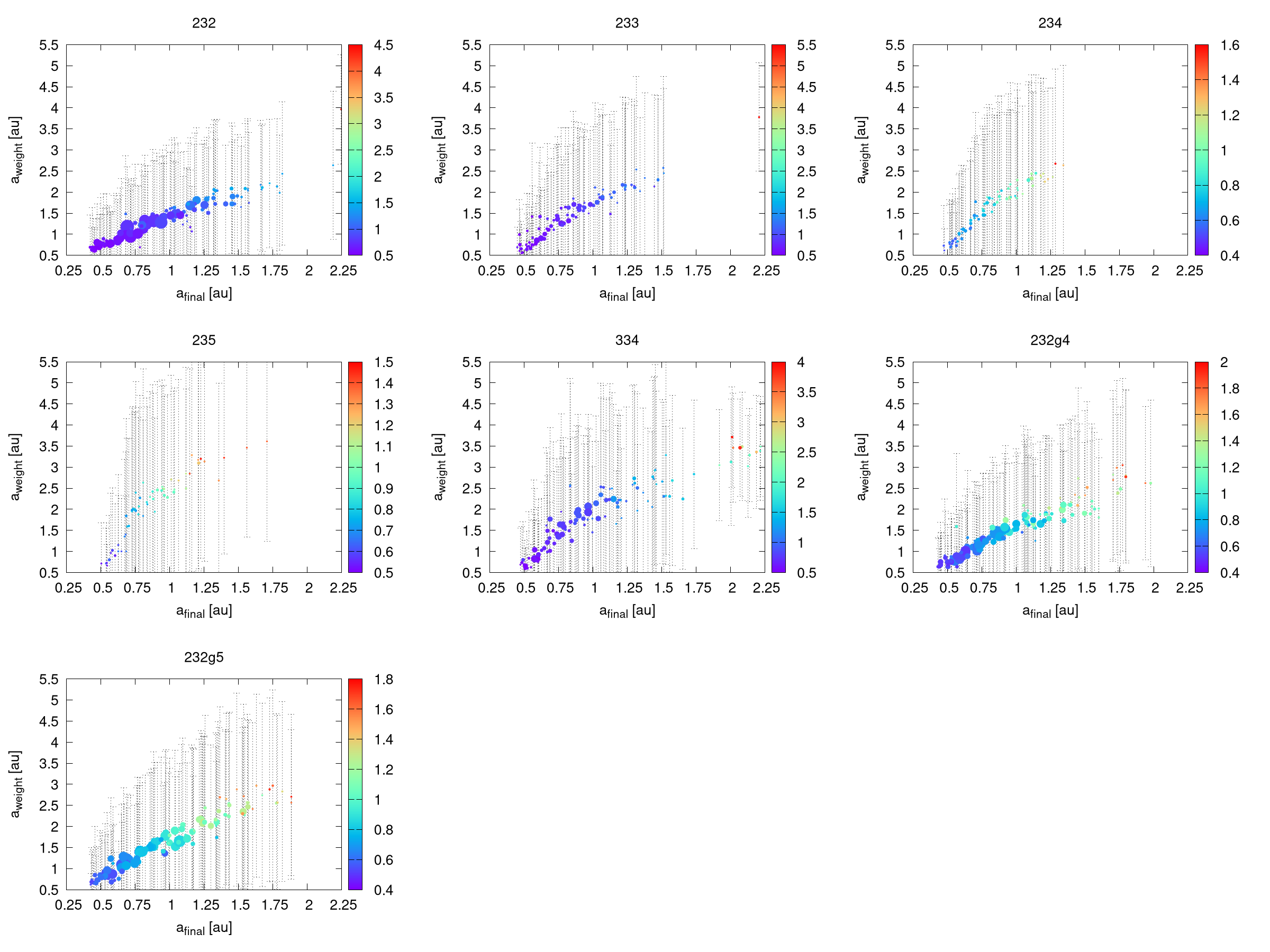}}
\caption{This figure shows the approximate feeding width of the embryos, depicting the final semi-major axis ($x$) versus the mass-weighted mean semi-major axis of the accreted material ($y$). The vertical error bars show the standard deviation/range of the feeding zones. There is clear correlation between the final and mass-weighted semi-major axis, showing that accretion is still mostly local but with a large random component. The colour coding indicates the initial semi-major axis of the embryo. For some small embryos this value is large because they came from beyond the composition changeover distance and went through a few mergers with material from the outer region.}
\label{fig:accrrange}%
\end{figure*}

\begin{figure*}
\resizebox{\hsize}{!}{\includegraphics{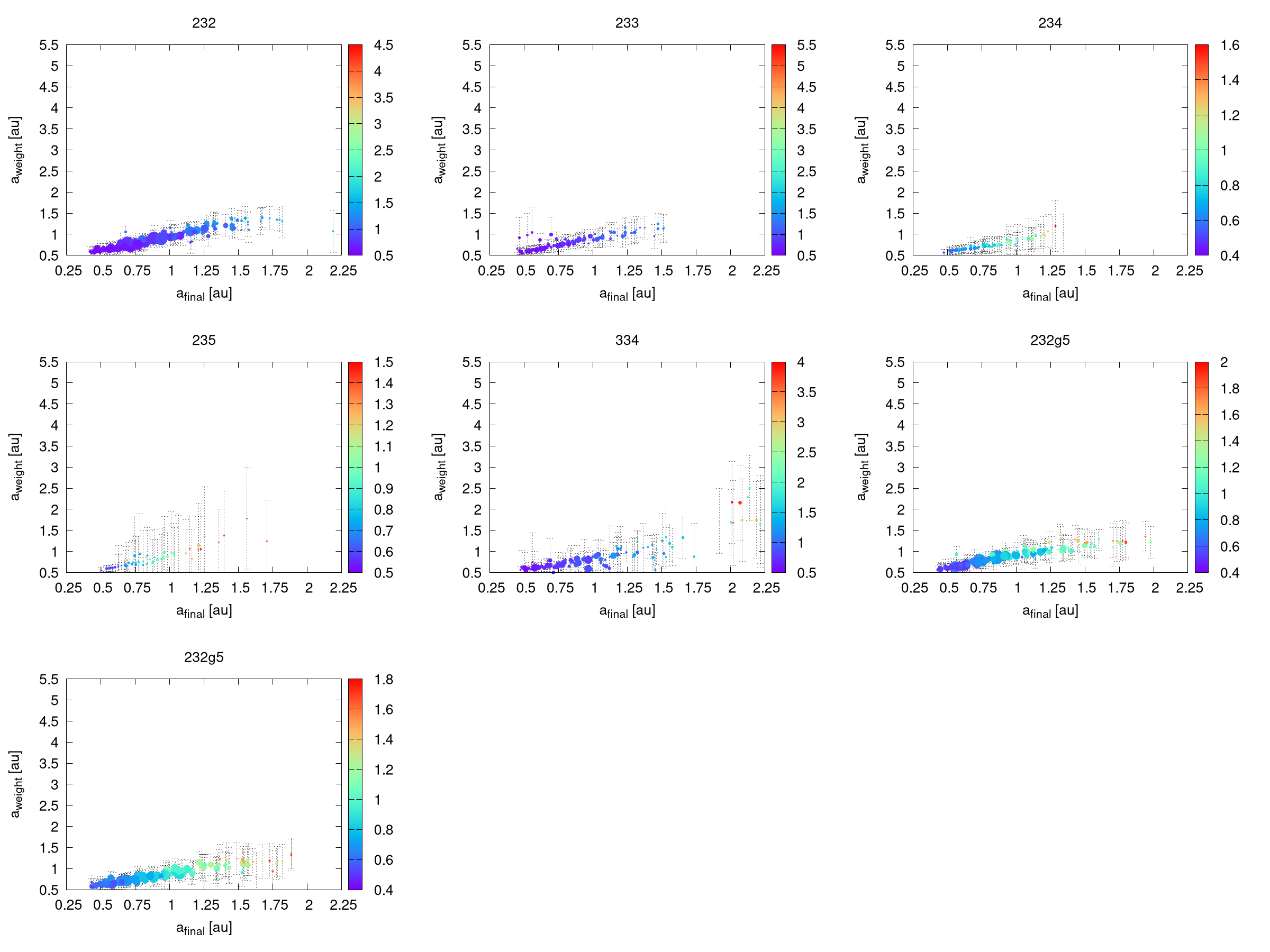}}
\caption{This is the same as Fig.~\ref{fig:accrrange} but now I ignore the mass accreted from {\it jovian} group material. It is clear that the feeding zones are much narrower and the correlating slope is much shallower. This shows that incorporation of {\it jovian} group material plays an important role in final composition.}
\label{fig:accrrangenooss}%
\end{figure*}

The final mass-semimajor axis distribution of the embryos with mass $m \geq 0.01$~$M_\oplus$ of the different sets of simulations is depicted in Fig.~\ref{fig:comp}. The colour coding indicates the mass fraction, $f_{\rm T}$, of each body that comes from the {\it terrestrial} group; the mass fraction of {\it jovian} material is necessarily $f_{\rm J}=1-f_{\rm T}$. Some small embryos close to the lunar mass can have high $(>25\%)$ fractions of {\it jovian} material incorporated through collision(s) with {\it jovian} group planetesimals. What is immediately clear is that the implantation model creates a compositional gradient similar to what was found by \citet{RaymondIzidoro2017}: less {\it jovian} material reaches the inner regions of the Solar System than father out. The Mercury region closer than 0.5~au has up to 5\% {\it jovian} material while near the present orbit of Mars this fraction is typically 20\%.\\

An additional trend is also clear. The top three panels and the middle-left panel all begin with 2~$M_\oplus$ of terrestrial group material spread over 1.5 au, 2.5 au, 3.5 au and 4.5 au. The 3~$M_\oplus$ of jovian material is spread across 6.5 au, 5.5 au, 4.5 au and 3.5 au respectively. The decreasing surface density of the {\it terrestrial} group material necessarily implies that the surface density of solids decreases as the width of its annulus decreases, which automatically explains why the top-left panel has the most massive embryos and the middle-left panel has the least massive embryos. Clearly the implantation of {\it jovian} group material cannot compensate for the decreasing surface density of the {\it terrestrial} group material. However, the distribution of {\it jovian} group material is different between the top-left and middle-left panels: apart from a few bodies near 2.25~au the mass fraction of {\it jovian} group material absorbed by the embryos with $a<2$~au is $f_{\rm J} \lesssim 15\%$. In contrast, in the middle-left panel embryos near 1.5~au have $f_{\rm J} \sim 35\%$. The middle-right and bottom panel both have the {\it terrestrial} group material truncated at 2 au so that these cases also have fast embryo growth. Last, several simulations have stranded embryos beyond 2~au which are almost entirely of {\it jovian} group material. Subsequent evolution will reveal where these bodies end up. \\

The same final outcomes of some of the the simulations is shown in Fig.~\ref{fig:snapcomp}, this time depicting the eccentricities versus semi-major axis of planets, and also of the planetesimals. For this figure, every body with a mass $m\geq0.01$~$M_\oplus$ is considered an embryo, and everything else a planetesimal. The colour coding is the same as in the previous figure and the size of the symbols is proportional to a body's mass. Red dots indicate planetesimals from the assumed {\it jovian} region while black planetesimals originated from the assumed {\it terrestrial} region. It is clear that the planetesimals are thoroughly mixed, apart from in the top panel where the planetesimals with $a>2$~au are predominantly red. This has implications for the composition of late accretion impactors on the terrestrial planets after the Moon-forming event.\\

For each embryo I tracked their accretion histories and computed their feeding zones. This is a proxy for the region in the protoplanetary disc from which the embryos sample most of their mass. The feeding zone of an embryo is quantified by the mean $a_{\rm{feed}}$ and width $\sigma_{\rm{feed}}$ of the initial semi-major axes of the solids accreted by it throughout its growth history weighted by their masses \citep{KaibCowan2015}. These quantities are expressed mathematically as \citep{Woo2018}
\begin{equation}
a_{\rm{feed}} = \frac{ \sum_{i}^{N} m_i a_i }{ \sum_{i}^{N} m_i },\ \rm{and}
\end{equation}

\begin{equation}
\sigma_{\rm{feed}}^2 = \frac{ \sum_{i}^{N} m_i (a_i - a_{\rm{feed}})^2}{\frac{N-1}{N} \sum_{i}^{N} m_i },
\end{equation}
where $m_i$ and $a_i$ are the mass and semi-major axis of the \textit{i}th body accreted by the embryo, and $N$ is the total number of bodies accreted. Together, $(a_{\rm feed }\pm\sigma_{{\rm feed}})$ define the feeding zone, although in practice this region is not symmetric.\\

In Fig.~\ref{fig:accrrange} I plot the final semi-major axis of the embryos at 5~Myr of evolution versus their mass-weight semi-major axis computed above. The colour coding indicates their original semi-major axis, symbol sizes are proportional to the mass of a body and the error bars are the $\sigma_{\rm feed}$ values. The colour coding indicates that, apart from the middle-middle panel, most embryos began as planetesimals with semi-major axis $a_{\rm init}<2$~au because that is where most of the mass in planetesimals is initially concentrated. Each panel shows a linear correlation between $a_{\rm final}$ and $a_{\rm feed}$. This is in contrast to the results by \citet{Woo2021}, who found no correlation. However, \citet{Woo2021} explain the lack of such a correlation due to the migration of embryos. Once they removed the effects of migration, or when the gas giants are assumed to reside on circular orbits, the correlation is obtained. In the top row the slope of the linear correlation increases: in the top-left panel the value of $a_{\rm feed}\sim 2$ when $a_{\rm final}\sim 1.25$, while this increases to $a_{\rm feed}\sim 3$ in the middle-left panel. This agrees with the result of Fig.~\ref{fig:comp} where the middle-left panel appears to have more m{\it jovian} material implanted than the top-left panel. Thus, even though the embryos become smaller due to the decreasing surface density of {\it terrestrial} material, they absorb a greater fraction of {\it jovian} material as their more massive counterparts in the simulations of the top panel, so that their mass-weighted semi-major axes are skewed higher.\\

To drive this point home I repeated the same calculation but I ignored all mass incorprated from the {\it jovian} group. The result is displayed in Fig.~\ref{fig:accrrangenooss}. Once again a correlation is visible, but the slope is much lower and $a_{\rm feed}$ never exceeds 2~au and the feeding zones are much reduced. The width of the feeding zones are comparable to those of the circular gas giants (CJS) and no gas disc cases of \citet{Woo2021} i.e. $\sigma_{\rm feed} \sim 0.25$--0.5~au. \\

I can quantify the amount of mass in embryos, and in planetesimals that consist of {\it terrestrial} and {\it jovian} group material, respectively. I further list the total mass in embryos that comes from the {\it jovian} reservoir and the average embryo mass in Table~\ref{tab:massdist}. The average fraction of {\it jovian} material incorporated in the embryos ranges from 6.5\% to 22\% while the fraction of {\it jovian} group material ending up in the inner solar system ranges from 22\% to 60\% depending, decreasing as the boundary us moved farther out. My implantation efficiency is about 20\%-25\% for material coming from beyond Jupiter, which is comparable but slightly larger than the value found by \citet{RaymondIzidoro2017}. \\

\begin{figure}
\resizebox{\hsize}{!}{\includegraphics{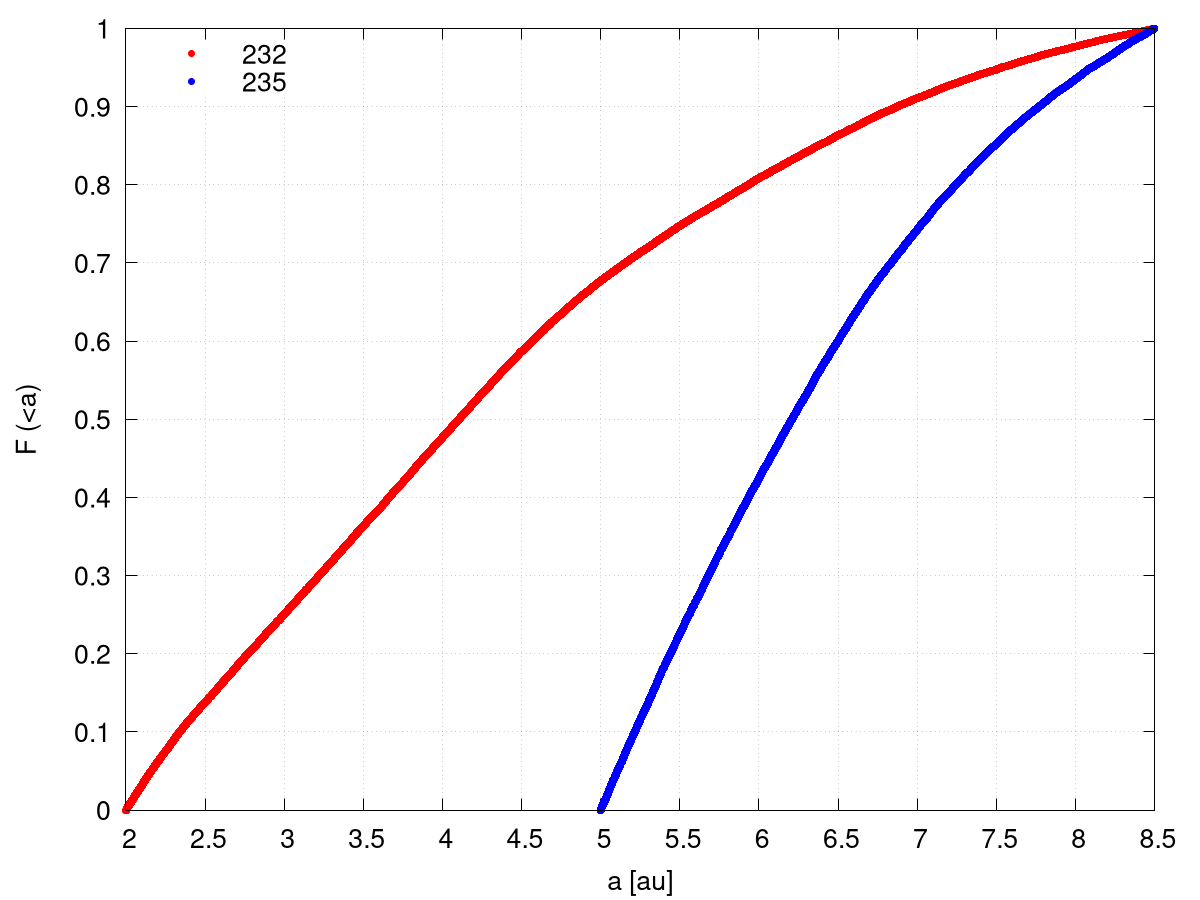}}
\caption{Cumulative distribution of the initial semi-major axis of {\it jovian} group material that ends up in the inner solar system, for the 232 and 235 sets of simulations. Between 10\% to 25\% of implanted material could be from the Saturn region, depending on where the changeover occurred.}
\label{fig:cumaj}%
\end{figure}

It has been suggested that the group of carbonaceous chondrite meteorites span a range of formation distances in the outer Solar System \citep[e.g.][]{VanKooten2016}, with the CO, CM and CV group forming closer to the Sun than CB, CH and CR \citep{Marrocchi2018,Marrocchi2022} and the CI the farthest away \citep{Alexander2013}. Although highly uncertain, the CO-CM-CV group could have originated near Jupiter \citep{Marrocchi2018} and the CI farther out \citep{Krot2015}. This could suggest that the the CB-CH-CR group formed near Saturn, and thus it is interesting to trace how much material from Saturn's region enters the inner solar system. The approximate dynamical range of the giant planets is 3.5 Hill radii, so that Jupiter's dynamical range is 4~au to 7~au, and Saturn's is 7~au to 8~au. I thus consider material with original semi-major axis $a>7$~au come from the Saturn region, and I compute the fraction of {\it jovian} group material that ends up into the inner Solar System that comes from Saturn's region. Figure~\ref{fig:cumaj} shows the cumulative distribution of the initial semi-major axis of {\it jovian} group material that was implanted into the inner solar system for the 232 and 235 sets of simulations; these two span the range of possibilities. From the figure I conclude that from 9\% to 26\% of {\it jovian} group material ends up in the inner solar system, depending on where the boundary between the two groups is. In other words, of the {\it jovian} group material that makes it into the inner Solar System, on average 15\% comes from the Saturn region. From Table~\ref{tab:massdist} I compute the total amount of mass in planetesimals originating from the Saturn region to be 0.1--0.25~$M_\oplus$.\\

The final thing I report on is that typically the fraction of planetesimals implanted beyond Saturn is $10^{-4}$. This is much lower than that of \citet{RaymondIzidoro2017}. However, their implaneted planetesimals beyond Saturn were all 100~km in diameter, or smaller. None of my planetesimals are smaller than 600~km in diameter, and indeed \citet{RaymondIzidoro2017} report almost no planetesimals being parked on stable orbits beyond Saturn when their planetesimal diameter is 1000~km. Thus I consider my results to be in agreement with theirs.

\begin{table}
\begin{tabular}{r|cccccc}
Set & $m_{\rm tot}$ & $f_{\rm emb}$ & $f_{\rm pl,J}$ & $f_{\rm emb,J}$ & $f_{a>7}$ & $\langle m_{\rm emb} \rangle$\\ \hline \\
232 & 3.96 & 20 & 18 & 16 & 8.9 & 0.055\\
233 & 3.90 & 27 & 43 & 6.5 & 12 & 0.030\\
234 & 3.65 & 16 & 50 & 34 & 17 & 0.024\\
235 & 3.43 & 12 & 62 & 26 & 26 & 0.021\\
334 & 2.88 & 11 & 66 & 23 & 16 & 0.029\\
232g4 & 3.35 & 33 & 25 & 9.0 & 14 & 0.049\\
232g5 & 2.94 & 38 & 23 & 11 & 22 & 0.055
\end{tabular}
\caption{The final mass the simulations in the inner solar system. The first column is the total remaining mass in emryos and planetesimals (in $M_\oplus)$ $f_{\rm emb}=m_{\rm emb}/m_{\rm tot}$ is the average fraction of total mass in embryos, $f_{\rm pl,J}=m_{\rm pl,J}/m_{\rm tot}$ is the same for {\it jovian} group planetesimals, $f_{\rm emb,J}=m_{\rm emb,J}/m_{\rm emb,tot}$ is the average mass fraction of {\it jovian} group material incorporated into the embryos, and $f_{a>7}$ is the fraction of planetesimals in the inner solar system that come from beyond 7~au. The final column is the average individual embryo mass.}
\label{tab:massdist}
\end{table}

\section{Discussion}
The study performed here is a direct continuation of the studies initiated by \citet{RaymondIzidoro2017}, \citet{Woo2021} and \citet{Lau2024,Raorane2024}. This work is an attempt to trace the growth of the terrestrial planets in the presence of the growing gas giants. I go beyond the results of \citet{RaymondIzidoro2017} by following the planetesimals scattered inwards by Jupiter and Saturn into planetary embryos and, ultimately, into the terrestrial planets in a forthcoming publication. Since this is a first attempt specific choices were made, which I have outlined above. As always, there is room for improvement if given more time and resources.\\

First, the growth rate of Jupiter adopted here is very fast, though comparable to what was adopted by \citet{RaymondIzidoro2017}. Yet the pebble accretion simulations of \citet{Lau2024, Raorane2024} show that it takes about 1~Myr to form Jupiter's core, and 3~Myr for Saturn's. One may debate the validity of these results, but \citet{Raorane2024} shows that the giant planets likely formed sequentially, otherwise Saturn would not have stopped growing at 95~$M_\oplus$. It is likely that the cosmochemical dichtomy existed by the time the Solar System was 1-2~Myr old \citep{Kruijer2017}, and irrespective of the mechanism that created it, it is difficult to argue that Jupiter formed earlier than this. As such, a repetition of this study with a slower growth of Jupiter is warranted, and is ongoing.\\

Second, I have kept the mass of planetesimals in the outer disc at 3~$M_\oplus$ and that of the inner disc is mostly 2~$M_\oplus$. The choice for the inner disc was motivated by the total mass in the terrestrial planets, because I worked under the assumption that the total mass in the inner disc would be mostly preserved. However, spreading this mass over nearly 5~au resulted in a very slow embryo growth, inconsistent with the growth timescale of Mars. Only when the inner disc is truncated at 2~au, and perhaps at 3~au, do I produce Mars-sized embryos within 5~Myr. This may point to a very low surface density existing in the region of the main asteroid belt at the time that the gas giants formed, but further studies need to rule this in or out. \\

Third, the high implantation efficiency implies that the fraction of embryo mass with {\it jovian} group material composition is generally more than 10\%, seemingly inconsistent with cosmochemical Monte Carlo mixing models \citep{Fitoussi2016,Dauphas2017,Dauphas2024}. This problem could be remedied by lowering the mass of the outer disc to 2~$M_\oplus$ or even 1~$M_\oplus$. Simulations with a 1~$M_\oplus$ outer disc are underway.\\

Fourth, a more contentious issue is the role of the $\nu_5$ resonance. \citet{Woo2021} discussed at length that the EJS scenario, in which the resonance operated, was a problem because it and the planetary migration caused the planets to all have the same feeding zones. In this study the feeding zones show a clear trend that $a_{\rm weight}$ increases with the final semi-major axis -- with and without accounting for the outer disc planetesimals. Thus it appears that the feeding zone problem that \citet{Woo2021} ran into was more the result of migration rather than the sweeping resonance. Still, a validation of this result with circular Jupiter and Saturn would be warranted.\\

Fifth, I did not include planet migration for reasons stated earlier: it would cause too much mass to enter the Mercury region of the Solar System, and because of needing to frequently adjust the $m_{\rm Giant}$ parameter while the simulations were running. However, a tentative case could be made that migration is not a huge problem if the terrestrial planets formed from a ring \citep{Woo2023}, or if there was a pressure maximum in the disc near Venus that caused outward migration in the Mercury region \citet{Clement2021a}. A more detailed study with a disc that has a pressure maximum near Venus and one near Jupiter is reserved for future work.\\

\section{Conclusions}
I have run numerical N-body simulations of TPF when the gas giants were growing. My main conclusions with the limitations of my study are the following. 

\begin{itemize}
    \item Between 22\% and 60\% of material that is considered isotopically to be of {\it jovian} group composition is implanted in the inner Solar System, depending on where the boundary between the two groups of material is located. Under the assumption that there was initially 3~$M_\oplus$ of this material, the implanted mass ranges from 0.66 to 1.8~$M_\oplus$.
    \item The implantation efficiency of material originating from beyond Jupiter is about 20\%, comparable to that found by \citet{RaymondIzidoro2017}.
    \item Between 6\% and 22\% of {\it jovian} group material is incorporated into planetary embryos after 5 Myr. This appears higher than what is suggested from cosmochemical experiments, but can be remedied by lowering the mass of the outer disc.
    \item Embryos have a compositional gradient, with {\it jovian} group material increasing with increasing distance.
    \item Between 0.1 to 0.25~$M_\oplus$ implanted in the inner solar system could have come from the Saturn region. This material could be the source population of the CB, CH and CR carbonaceous chondrites.
    \item Only inner discs truncated at 2~au and possibly at 3~au produce Mars-mass embryos within 5~Myr. This suggests that the current region of the asteroid belt could have had a low initial mass when the Solar System formed.
    \item Remaining planetesimals have a completely mixed composition, which eventually has implications for late accretion. 
\end{itemize}
Future simulations exposing the remaining planetesimals and embryos to late giant planet migration \citep{Tsiganis2005} and dynamical evolution will reveal whether this scenario is consistent with the isotopic composition of the terrestrial planets.

\section*{Acknowledgments}
I am thankful to Jason Woo for valuable feedback. I gratefully acknowledge the EuroHPC Joint Undertaking for awarding me access to Vega at IZUM, Slovenia, project EHPC-REG-2022R03-047, and MeluXina at LuxProvide, Luxembourg, project EHPC-REG-2023R03-050. I also acknowledge PRACE for awarding access to the Fenix Infrastructure resources at JUSUF-HPC, Germany, which are partially funded from the European Union’s Horizon 2020 research and innovation programme through the ICEI project under the grant agreement No. 800858. This study is supported by the Research Council of Norway through its Centres of Excellence funding scheme, project No. 332523 PHAB. GENGA can be obtained from \url{https://bitbucket.org/sigrimm/genga/}

\bibliographystyle{aa}
\bibliography{main}
\end{document}